\documentclass[rnote]{aa} 
\usepackage[utf8]{inputenc}
\usepackage[T1]{fontenc}
\usepackage{graphicx}
\usepackage{txfonts}
\usepackage{amsfonts}
\usepackage{natbib}
\usepackage{float}
\usepackage{lscape}
\usepackage{nicefrac}

\begin{document}
   \title{Far-infrared CO and H$_2$O emission in intermediate-mass protostars}

   \author{M.~Matuszak\inst{1}, A.~Karska\inst{1}, L.~E.~Kristensen\inst{2},
   G.~J.~Herczeg\inst{3}, Ł.~Tychoniec\inst{1}, T.~van Kempen\inst{4}, and A.~Fuente\inst{5}}
	\institute{
          	  $^{1}$ Astronomical Observatory, Adam Mickiewicz University, Słoneczna 36, PL-60-268 Poznań, Poland \\
          	  $^{2}$ Harvard-Smithsonian Center for Astrophysics, 60 Garden Street, 
          	  Cambridge, MA 02138, USA\\
          	  $^{3}$ Kavli Institut for Astronomy and Astrophysics, Yi He Yuan Lu 5, HaiDian Qu, Peking University, Beijing,
          	  100871, PR China\\
   			  $^{4}$ Leiden Observatory, Leiden University, P.O. Box 9513,
          	  2300 RA Leiden, The Netherlands\\
          	  $^{5}$ Observatorio Astron\'{o}mico Nacional (OAN, IGN), Apdo 112, 28803 Alcal\'{a} de Henares, Spain \\
    \email{agata.karska@amu.edu.pl}
             }

   \date{Received March 4, 2015; accepted April 13, 2015}
	\titlerunning{Intermediate-mass YSOs with PACS}
	\authorrunning{M.~Matuszak et al. 2015}

  \abstract
  {Intermediate-mass young stellar objects (YSOs) provide a link 
  to understand how feedback from shocks and UV radiation scales from low to high-mass star forming regions.}
  {Our aim is to analyze excitation of CO and H$_2$O in deeply-embedded 
  intermediate-mass YSOs and compare with low-mass and high-mass 
  YSOs.}
  {Herschel/PACS spectral maps are analyzed for 6 YSOs with bolometric
   luminosities of $L_\mathrm{bol}\sim10^2 - 10^3$ $L_\odot$. The maps cover 
   spatial scales of $\sim 10^4$ AU in several CO and H$_2$O lines located in the $\sim55-210$ $\mu$m range.}
  {Rotational diagrams of CO show two temperature components at $T_\mathrm{rot}\sim320$ K and
  $T_\mathrm{rot}\sim700-800$ K, comparable to low- and high-mass protostars probed at similar spatial scales. 
  The diagrams for H$_2$O show a single component at $T_\mathrm{rot}\sim130$ K, as seen in 
  low-mass protostars, and about $100$ K lower than in high-mass protostars. Since the 
  uncertainties in $T_\mathrm{rot}$ are of the same order as the difference between the intermediate and high-mass 
  protostars, we cannot conclude whether the change in rotational temperature occurs at a specific 
  luminosity, or whether the change is more gradual from low- to high-mass YSOs.}
  {Molecular excitation in intermediate-mass protostars is comparable to the central 
  $10^{3}$ AU of low-mass protostars and consistent within the uncertainties with the 
  high-mass protostars probed at $3\cdot10^{3}$ AU scales, suggesting similar shock 
  conditions in all those sources.}
  

   \maketitle
\section{Introduction}
Feedback processes associated with the collapse of protostellar envelopes 
at $10^3-10^4$ AU scales limit the accretion onto the protostar and contribute 
to the overall low efficiency of transferring gas into stars on global 
scales \citep{Of09,Krum14}. Calculating the physical conditions 
help to identify the most relevant phenomena 
 and constrain their role in the star formation process \citep{Ev99}. 
Since gas in protostellar envelopes is heated to temperatures much higher than the dust temperatures,
molecular transitions are the suitable tracers of physical conditions of hot ($T\gtrsim100$ K) gas
 around protostars. In particular, the far-infrared (IR) 
lines of CO and H$_2$O dominate the cooling of hot and dense gas \citep{GL78}. 
The excitation of CO and H$_2$O depends on the local physical conditions (temperature, density) and thus 
is crucial to determine which physical mechanisms are responsible for the gas heating and
to study whether the energetics involved in the feedback 
scale from low- to intermediate- to high-mass young stellar objects (YSOs).

Recent observations of CO and H$_2$O lines with the Photodetector 
Array Camera and Spectrometer \citep[PACS,][]{Po10} on board \textit{Herschel} 
found evidence for large columns of dense ($\gtrsim 10^4$ cm$^{-3}$) and hot ($\gtrsim 300$ K) 
gas towards low-mass ($L_\mathrm{bol}\lesssim10^2 L_\odot$) protostars \citep{vK10,He12,Ma12,Ka13,Gr13,Li14}, originate largely
 from UV-irradiated shocks associated with jets and winds \citep{Ka14b}.
CO and H$_2$O line luminosities of the high-mass protostars (with $L_\mathrm{bol}\sim10^{4}-10^{6}$ $L_{\odot}$)
follow the correlations with bolometric luminosities found in the low-mass protostars \citep{Ka14} and 
show similar velocity-resolved line profiles regardless of the mass of the protostar \citep{Yi13,IreneCO,vT13}. 
In contrast, rotational temperatures of H$_2$O are lower and the H$_2$O fraction 
contributed to the total cooling in lines with respect to CO is higher for the low-mass protostars \citep{Ka14,Go15},
suggesting that the physical mechanism causing the excitation in low- and high-mass protostars are different.

\begin{figure*}
\sidecaption
\includegraphics[angle=90,width=12cm]{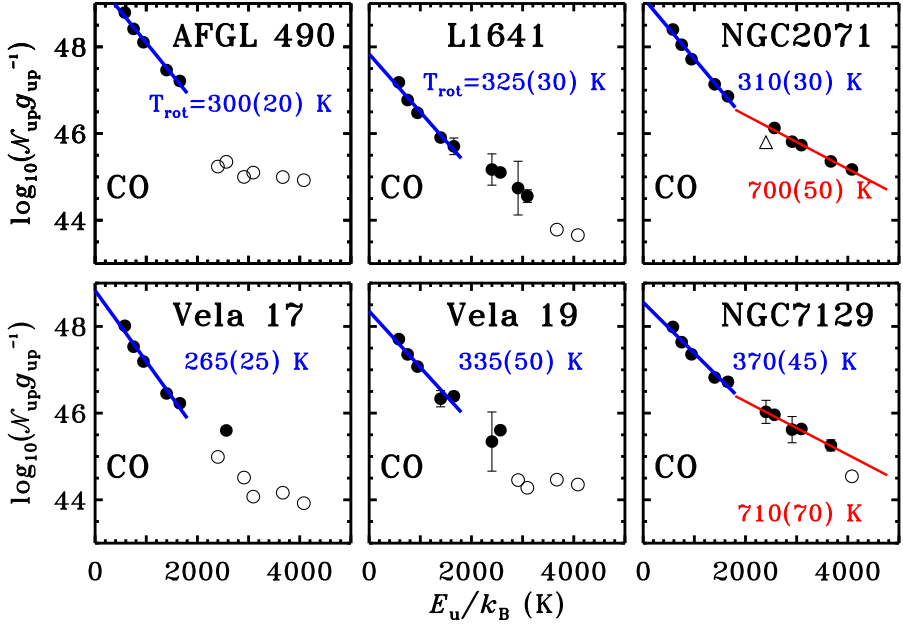}
\caption{\label{coex} Rotational diagrams of CO. 
 The base-10 logarithm of the number of
  emitting molecules from the upper level,
  $\mathcal{N}_\mathrm{up}$, divided by the degeneracy of the level,
  $g_\mathrm{up}$, is shown as a function of energy of the upper level
  in kelvins, $E_\mathrm{up}$. Detections are shown as filled
  circles, whereas three-sigma upper limits are shown as empty
  circles. Empty upper triangle corresponds to the line flux calculated using 
  a smaller area on the map than the rest of the lines. Blue lines show linear fits to the data and the
    corresponding rotational temperatures. Errors associated with the least-square linear fit are shown in brackets.}
\end{figure*}
\begin{figure*}[!tb]
\sidecaption
\includegraphics[angle=90,width=12cm]{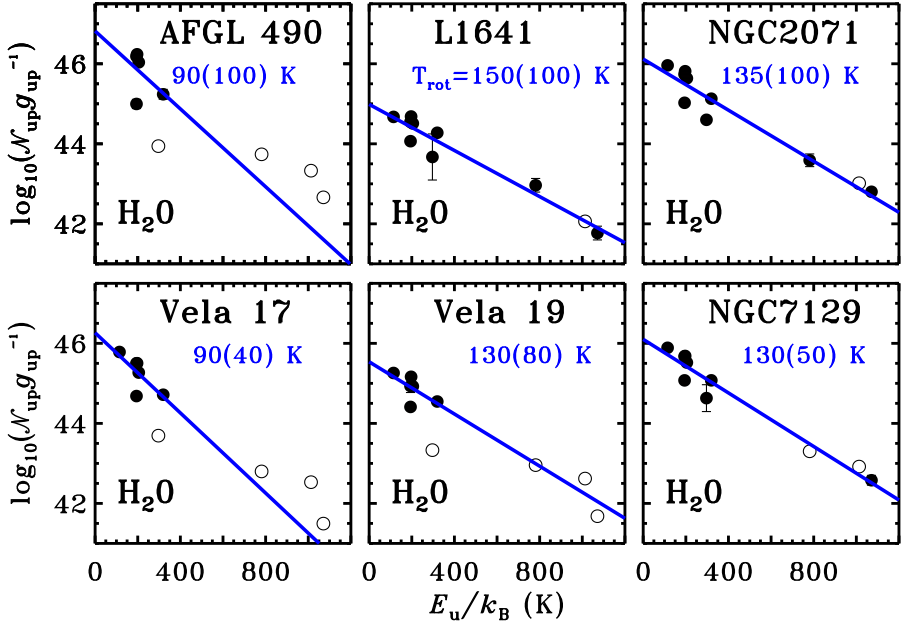}
\caption{\label{wex} Similar to Figure \ref{coex} but for H$_2$O.}
\end{figure*}
Intermediate-mass YSOs (with $L_\mathrm{bol}\sim10^{2}-10^{3}$ $L_{\odot}$\footnote{
We use bolometric luminosity as a proxy of the 
protostellar mass for practical and historical reasons, but we note that $L_\mathrm{bol}$ changes
 significantly during the protostellar phase if the accretion is episodic \citep{Yo05,MD10}. Moreover,
 some of our intermediate-mass sources may be in fact a collection of unresolved low-mass protostars.}) 
provide a natural link between low- and high mass protostars, but their far-IR CO and H$_2$O emission has only 
been studied for a single protostar position of NGC 7129 FIRS 2 \citep{Fi10} and the outflow position of 
NGC 2071 \citep{Ne14}. CO emission alone was analyzed for two intermediate-mass protostars in Orion,
HOPS 288 and 370 \citep{Ma12}. In this paper, we present the analysis of PACS spectra for the 
full sample of intermediate-mass protostars from the `Water in star forming regions with Herschel' 
(WISH) key program \citep{WISH}, including the maps of NGC 7129 and NGC 2071 centered on the 
YSO position. These results complement the work by \citet{Wa13}, which describes the OH excitation in our source sample 
and the sample of low- and high-mass protostars for which CO and H$_2$O emission is discussed in \citet{Ka13,Ka14}.
The main question addressed is whether CO and H$_2$O rotational temperatures differ 
from low- to high-mass protostars. 

The paper is organized as follows. \S 2 briefly introduces the observations, 
\S 3 the excitation analysis using rotational diagrams, and \S 4 discusses the results.

\section{Observations}
\begin{table*}[tb!]
\caption{\label{tab:exc} CO and H$_2$O rotational excitation}
\centering
\renewcommand{\footnoterule}{}  
\begin{tabular}{lcccccccccccccc}
\hline \hline
Source & $D$ & $L_\mathrm{bol}$ & \multicolumn{2}{c}{Warm CO} & \multicolumn{2}{c}{Hot CO} & \multicolumn{2}{c}{H$_2$O} \\
~ & (pc) & ($L_{\odot}$) & $T_\mathrm{rot}$(K) & $\mathrm{log}_\mathrm{10}\mathcal{N}$ & $T_\mathrm{rot}$(K) & 
$\mathrm{log}_\mathrm{10}\mathcal{N}$ & $T_\mathrm{rot}$(K) & $\mathrm{log}_\mathrm{10}\mathcal{N}$ \\
\hline
AFGL 490  & 1000 & 2000 & 300(20) & 51.6(0.1) & \ldots & \ldots  & 90(100) & 48.3(1.2)  \\
L 1641 S3 MMS1   & 465 & 70 & 325(30) & 49.9(0.1) & \ldots & \ldots  & 150(100) &  46.8(0.1)   \\
NGC 2071  & 422 & 520 & 310(30) & 51.2(0.1) & 700(50) & 50.1(0.1)  & 135(100) & 47.9(0.1)    \\
Vela 17   & 700 & 715 & 265(25) & 50.8(0.2) & \ldots &  \ldots & 90(40) &  47.8(0.5)  \\
Vela 19   & 700  & 776  & 335(50) & 50.4(0.2) & \ldots &  \ldots  & 130(80) & 47.2(0.4)   \\
NGC 7129 FIRS 2 & 1250 & 430 & 370(45) & 50.7(0.2) & 710(70) & 49.9(0.2)  & 130(50)  & 47.8(0.1)   \\
\hline
\end{tabular}
\tablefoot{Distances and bolometric luminosities are taken from \citet{Wa13} and references therein.}
\end{table*}
Our sample includes 6 YSOs with bolometric luminosities from 70 to 2000 $L_\odot$ 
and located at an average distance of 700 pc (see Table \ref{tab:exc}). 
The sources were selected based on their small distances ($\lesssim$ 1 kpc) 
and location accessible for follow-up observations from the southern hemisphere 
\citep[for more details see \S 4.4.2. in][]{WISH}. Spectroscopy for all
sources was obtained with PACS 
as part of the WISH program. For observing details see Table \ref{log} in the Appendix.  

With PACS, we obtained single footprint spectral maps covering a field of view of $\sim47''\times47''$ and resolved 
into 5$\times$5 array of spatial pixels (spaxels) of $\sim9.4''\times9.4''$ each. At the distance to the sources,
the full array corresponds to spatial scales of $\sim 2-6\times 10^4$ AU in diameter, of the order of 
full maps of low-mass YSOs ($\sim 10^4$ AU) in \citet{Ka13} and the central spaxel spectra of more distant high-mass YSOs 
($\sim 3\times 10^4$ AU) in \citet{Ka14}.

The observations were taken in the line spectroscopy mode, which provide deep integrations of 
0.5--2 $\mu$m wide spectral regions within the $\sim$55--210 $\mu$m PACS range. Two nod positions 
were used for chopping 3$^\prime$ on each side of the source \citep[for 
details of our observing strategy and basic reduction methods, see][]{Ka13}. The full list of targeted 
CO and H$_2$O lines and the calculated line fluxes are shown in Table \ref{lines} in the Appendix. 
The quality of the spectra is illustrated in Figure \ref{spec} in the Appendix.
We note that the simultaneous non-detections of the H$_2$O 7$_{16}$-7$_{07}$ line at 84.7 $\mu$m 
and detections of the H$_2$O 8$_{18}$-7$_{07}$ line at 63.3 $\mu$m in L 1641, NGC 2071, and 
NGC 7129 FIRS 2 are related to the structure of the H$_2$O energy levels and not 
the differences in the sensitivities of the instrument in the 2nd and 3rd order observations. 
The 8$_{18}$-7$_{07}$ line is a backbone transition, with 
the Einstein A coefficient higher than the one for the 7$_{16}$-7$_{07}$ line.

The data reduction was done using HIPE v.13 with Calibration Tree 65 and subsequent 
analysis with customized IDL programs \citep[see e.g.][]{Ka14b}. The fluxes were calculated 
using the emission from the entire maps. Figure \ref{maps} in the Appendix illustrates 
that both the line and continuum emission peaks approximately at the source position, with 
small shifts in continuum due to mispointing. The extent of the line emission follows typically
the continuum pattern with the exception of Vela 17 where the line emission extends from NE to SW 
direction, while the continuum is centrally peaked. There is no contamination 
detected from the nearby sources or their outflows in the targeted regions.

\section{Rotational diagrams}
\subsection{Results}
\begin{figure}[!tb]
\begin{center}
\includegraphics[angle=90,height=7cm]{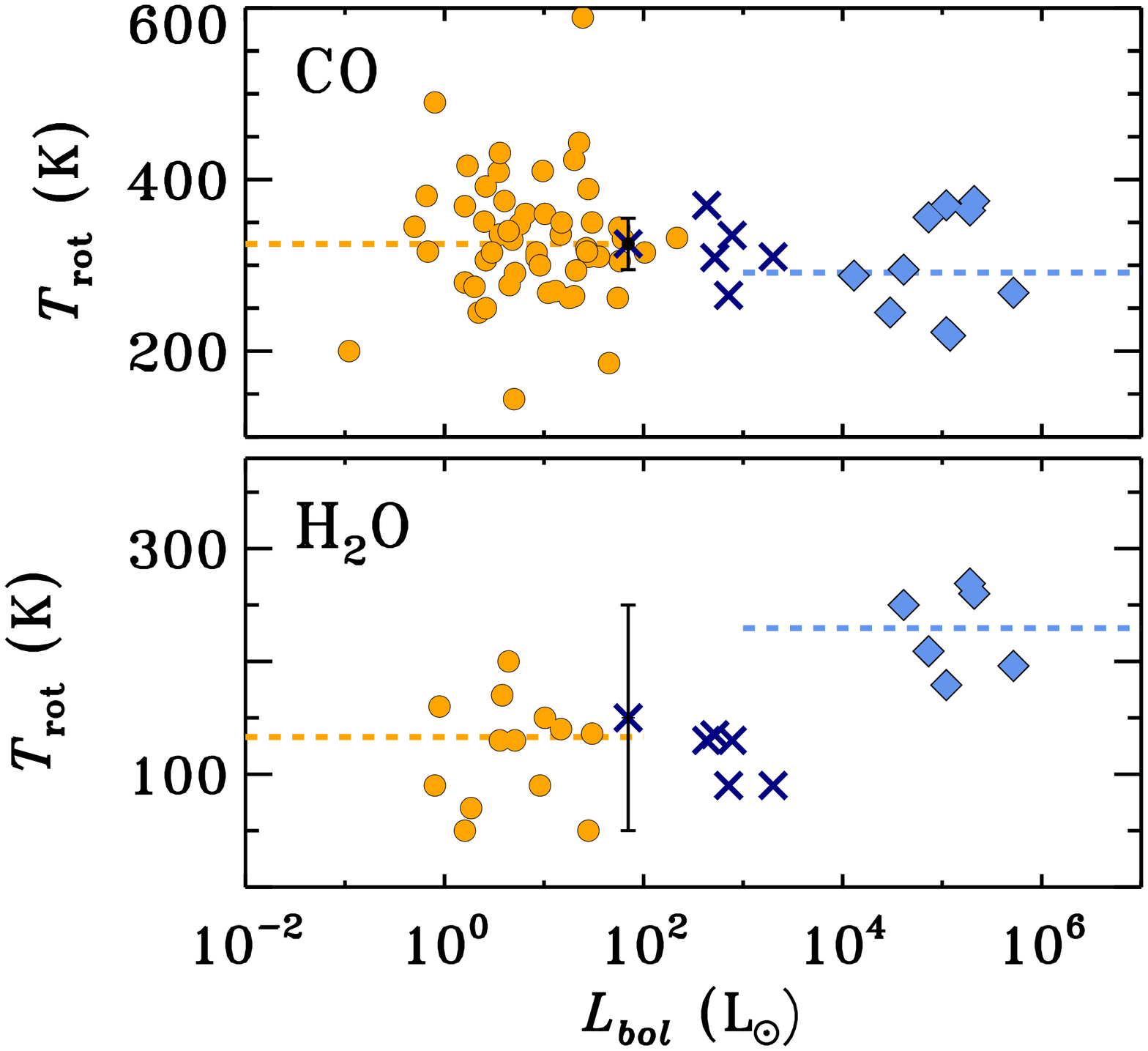}
\vspace{-0.4cm}

\caption{\label{trot} Rotational temperatures of \lq warm' ($E_\mathrm{up}=14-24$) CO (top) and H$_2$O (bottom) as function 
of bolometric luminosity. Orange circles show YSOs from 
\citet{Gr13}, \citet{Ka13}, and \citet{Ma12}. Dark blue crosses show intermediate-mass 
YSOs from the WISH program and light blue diamonds high-mass YSOs from \citet{Ka14}. 
A few sources with luminosities $\sim10^2$ $L_\odot$ shown in orange crosses are from
\citet{Ma12} and can be also regarded as intermediate-mass YSOs. Uncertainties 
of the CO and H$_2$O rotational temperatures are shown for L1641 S3 MMS1, which 
is representative of the sample.}
\end{center}
\end{figure}
Figure \ref{coex} shows rotational diagrams of CO calculated for all sources in the 
same way as in \citet{Ka14}. The corresponding rotational temperatures, $T_\mathrm{rot}$,  
and total numbers of emitting molecules, $\mathcal{N}$, are shown in Table \ref{tab:exc}.

All sources show a 300 K, \lq warm' CO component \citep{Ma12,Ka13,Gr13}, with a mean temperature 
of $\sim 320$ K for the range of total numbers of emitting molecules, $\mathcal{N}\sim 10^{50} - 10^{51}$. 
In addition, NGC2071 and NGC7129 show a \lq hot' CO component with temperatures of $\sim 700$ K 
and $\mathcal{N}\sim 10^{50}$. 

The rotational diagrams for H$_2$O, presented in Figure \ref{wex}, show a single component with 
a mean temperature of $\sim 120$ K. The scatter due to subthermal excitation and 
high opacities exceeds the uncertainties in the observed fluxes, similar to diagrams of low- and high-mass 
YSOs \citep[for the discussion of both effects see \S 4.2.2 of][]{Ka14}. 
The corresponding numbers of emitting molecules are about 3 orders of magnitude lower than 
for CO, $\mathcal{N}\sim 10^{47} - 10^{48}$.

\subsection{Comparison to low- and high-mass sources}

Figure \ref{trot} shows a comparison between rotational temperatures obtained for CO 
and H$_2$O for the intermediate-mass sources presented here and these quantities 
determined in the same way for low- and high-mass YSOs. 
The comparison is restricted only to the \lq warm' component 
seen in CO rotational diagrams due to the low number of sources with detections of the \lq hot' CO component; 
those will be discussed in detail in Karska et al. (in prep.). A bolometric luminosity 
is used here as a proxy for the protostellar mass, but in fact some of 
our intermediate-mass sources may be a collection of unresolved low-mass protostars.

The median $T_\mathrm{rot}$ of CO in low-mass protostars is 325 K, using the results from the WISH 
\citep{Ka13}, \lq Dust, Ice, and Gas in Time' \citep{Gr13}, and \lq Herschel Orion Protostar Survey' 
\citep{Ma12} programs for a total of about 50 sources. A comparable value of $\sim$290 K was 
found for 10 high-mass sources ($L_\mathrm{bol}\sim10^4 - 10^6$ $L_\odot$) in \citet{Ka14}.  
The range of CO $T_\mathrm{rot}$ of 265-370 K (see Table \ref{tab:exc}) determined 
here for intermediate-mass YSOs is thus fully consistent with previous results. The fact that the
 300 K CO component does not depend on the source bolometric luminosity 
over 6 orders of magnitude suggests the origin in a shock associated with the jet / winds 
impact on the envelope rather than in a photodissociation region where $T_\mathrm{rot}$
should scale with the UV flux and luminosity \citep{vK10,Vi12,Ma12,FP13,Kr13}. 

The median H$_2$O rotational temperatures (see the bottom panel of Figure \ref{trot}) 
for low- and high-mass YSOs are $\sim130$ K and $\sim230$ K, respectively \citep{Ka13,Ka14}. 
The range of temperatures obtained for the intermediate-mass YSOs (90-150 K, Table \ref{tab:exc}) 
is thus comparable to the low-mass sources. However, the uncertainties in the rotational
temperatures are high, of the order of $\sim100$ K, and do not account for the optical depth 
effects, the density effects, and the possible complexity of the line profiles (absorption and emission 
components). It is therefore unclear if there is
a true jump in rotational temperatures at $L_\mathrm{bol}\sim10^4$ $L_\odot$, or a smooth trend 
toward higher values of $T_\mathrm{rot}$. In either case, these higher excitation 
temperatures could be due to higher densities in the more massive envelopes.

\section{Summary}

We analyze the excitation of far-infrared CO and H$_2$O lines in 6 intermediate-mass 
YSOs observed in the WISH survey and compare the results to low- and high-mass protostars.
Rotational temperatures of CO and H$_2$O are found to be $\sim320$ K and $\sim120$ K, respectively,
and are consistent with low-mass and high-mass YSOs within the uncertainties. 
The large uncertainties in the H$_2$O rotational temperatures, of the order of 100 K, 
and the order of magnitude gap in the bolometric luminosity between intermediate- 
and high-mass protostars does not allow us to conclude whether the changes are a smooth 
function of luminosity. Still, the similarities in rotational 
temperatures seen for sources with luminosities spanning 6 orders of magnitude and 
probed at different spatial scales strongly suggest
the same excitation mechanism, the UV-irradiated shocks associated with jets and winds
for all sources across the luminosity range \citep{Kr13,Ka14b,Mo14}. 
 
\begin{acknowledgements}
The authors would like to thank the referee for the valuable comments 
which helped to improve the manuscript. Herschel is an ESA space observatory with science instruments provided
by European-led Principal Investigator consortia and with important
participation from NASA. AK acknowledges support from  
the Polish National Science Center grant 2013/11/N/ST9/00400. 
Research conducted within the scope of the HECOLS International Associated 
Laboratory, supported in part by the Polish NCN grant DEC-2013/08/M/ST9/00664.
\end{acknowledgements}   
   
\bibliographystyle{aa}
\bibliography{biblio14}

\appendix
\Online
\onecolumn

\section{Supplementary material}
\begin{figure}[!tb]
\begin{center}
\includegraphics[angle=90,height=12cm]{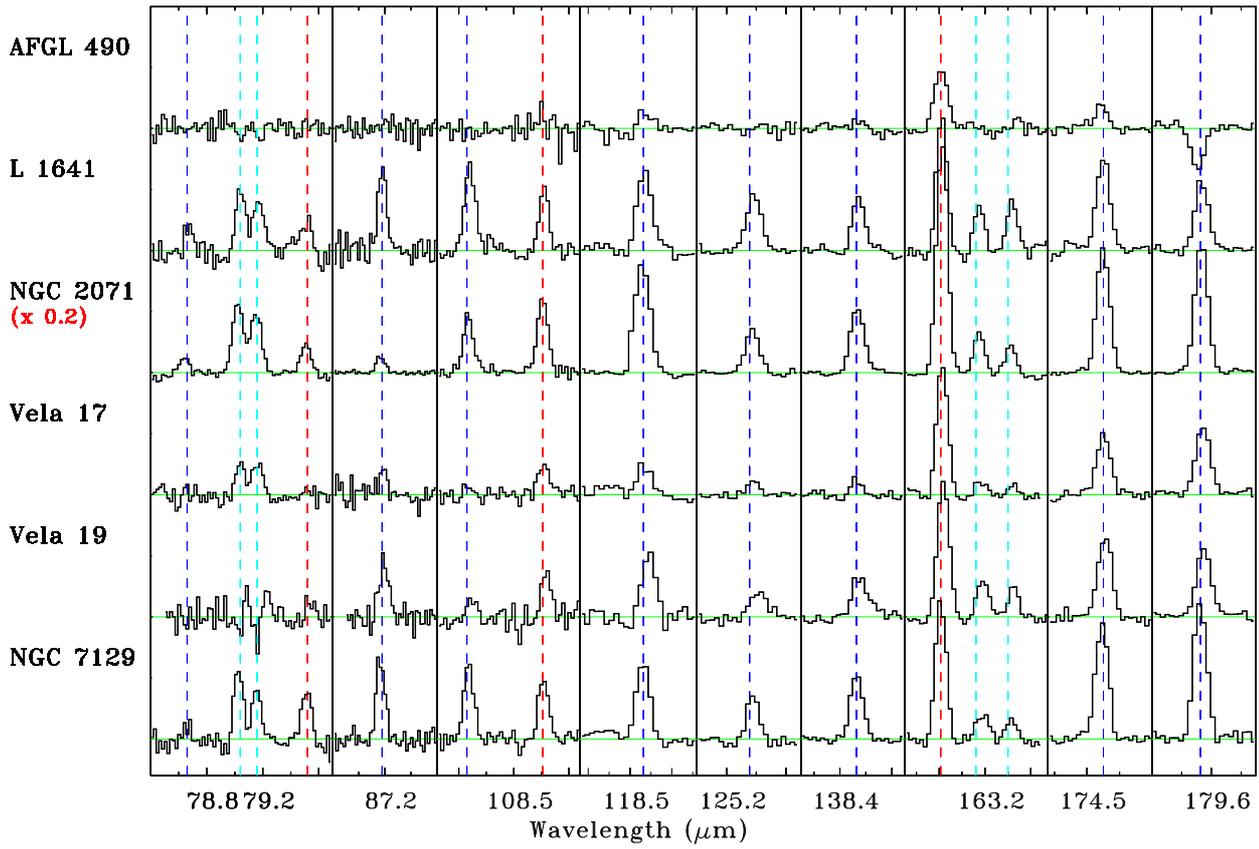}
\vspace{-0.4cm}

\caption{\label{spec} Spectral scans covering selected H$_2$O, CO and OH lines in the 
intermediate mass protostars from the WISH program. The rest wavelength of each line 
is indicated by dashed lines: blue for H$_2$O, red for CO, and light blue for OH. }
\end{center}
\end{figure}

\begin{figure*}[!tb]
  \begin{minipage}[t]{.5\textwidth}
  \begin{center}  
      \includegraphics[angle=90,height=8cm]{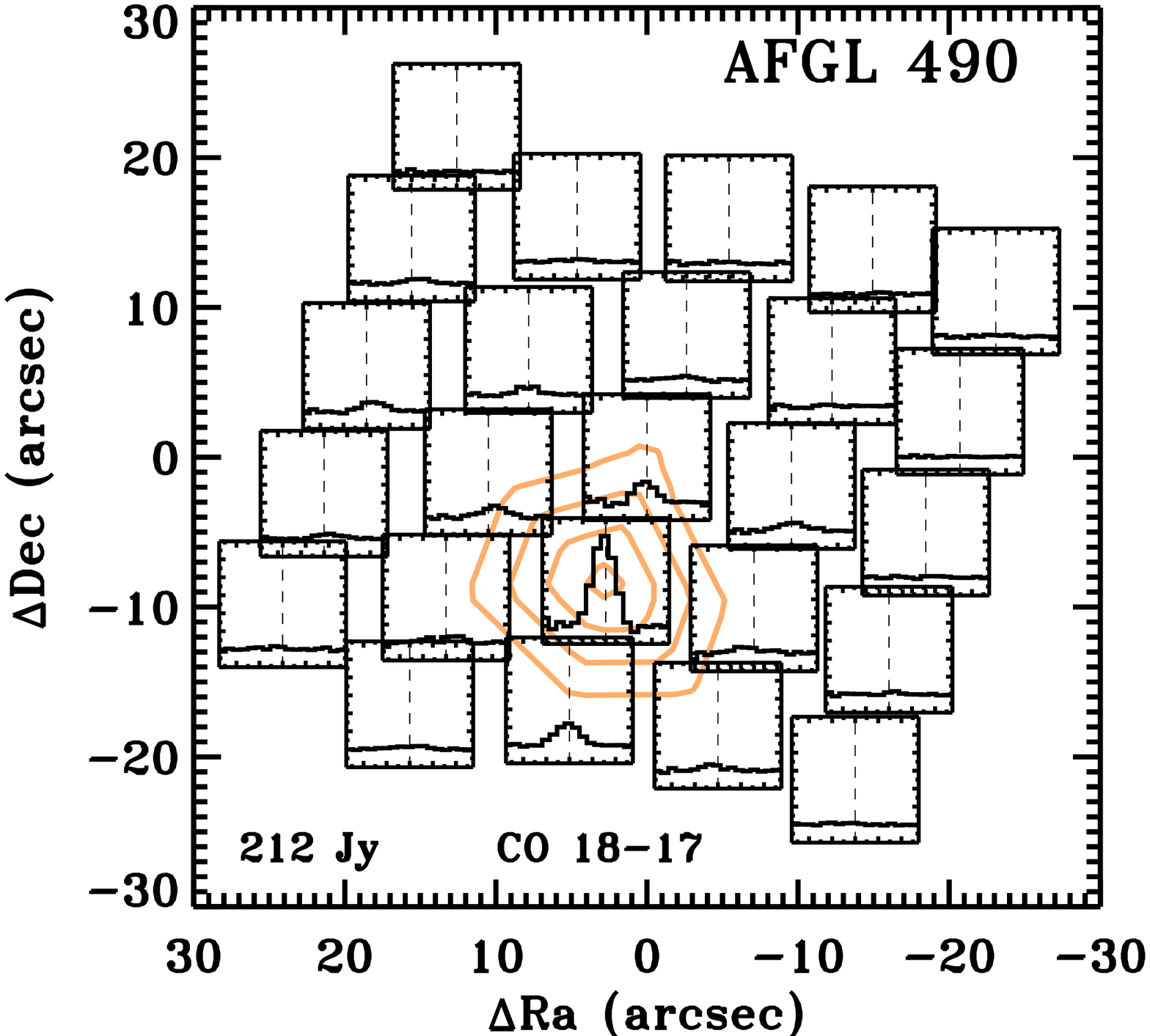}
               \vspace{+3ex}
       
      \includegraphics[angle=90,height=8cm]{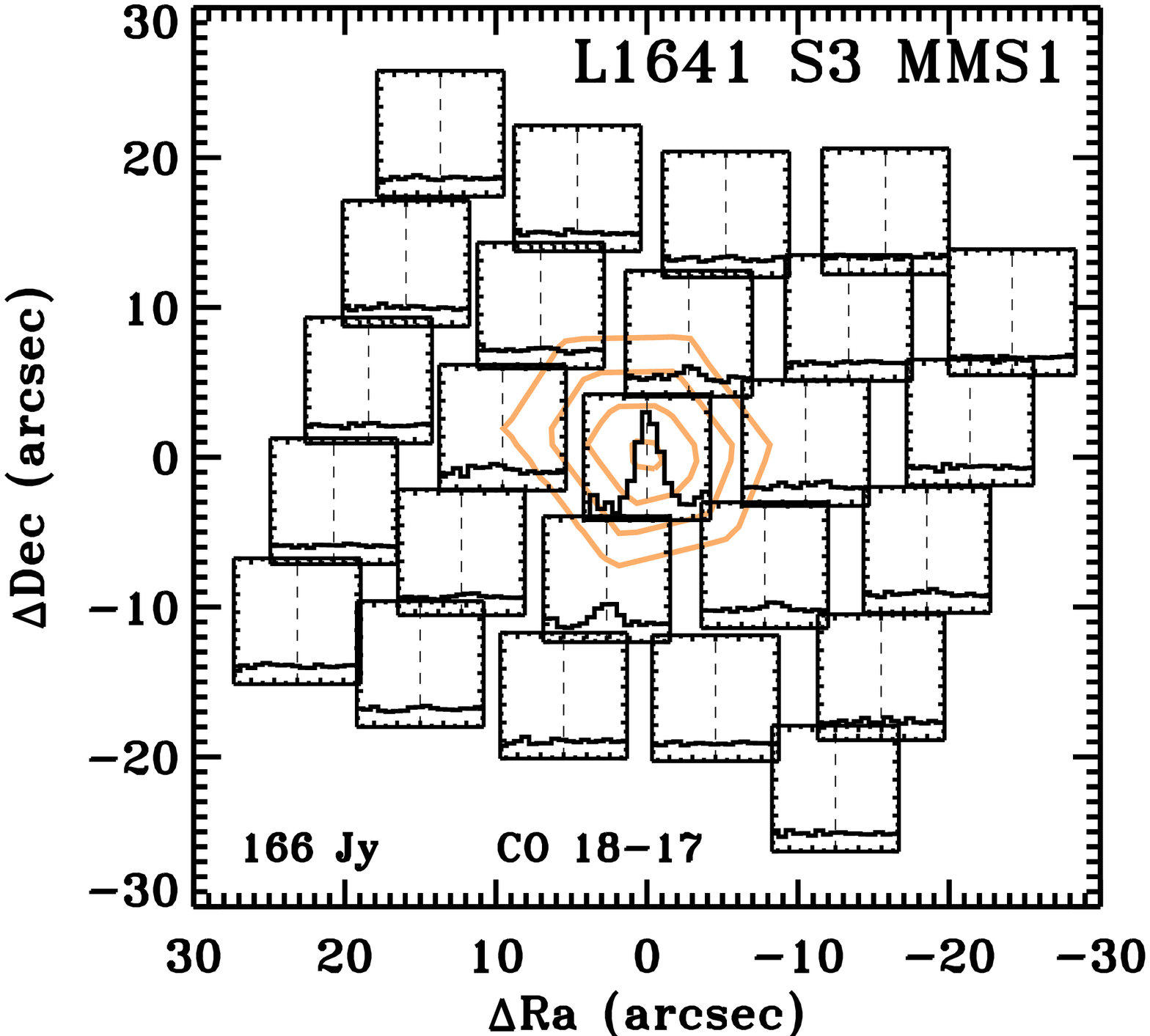}
  			    \vspace{+3ex}
            \includegraphics[angle=90,height=8cm]{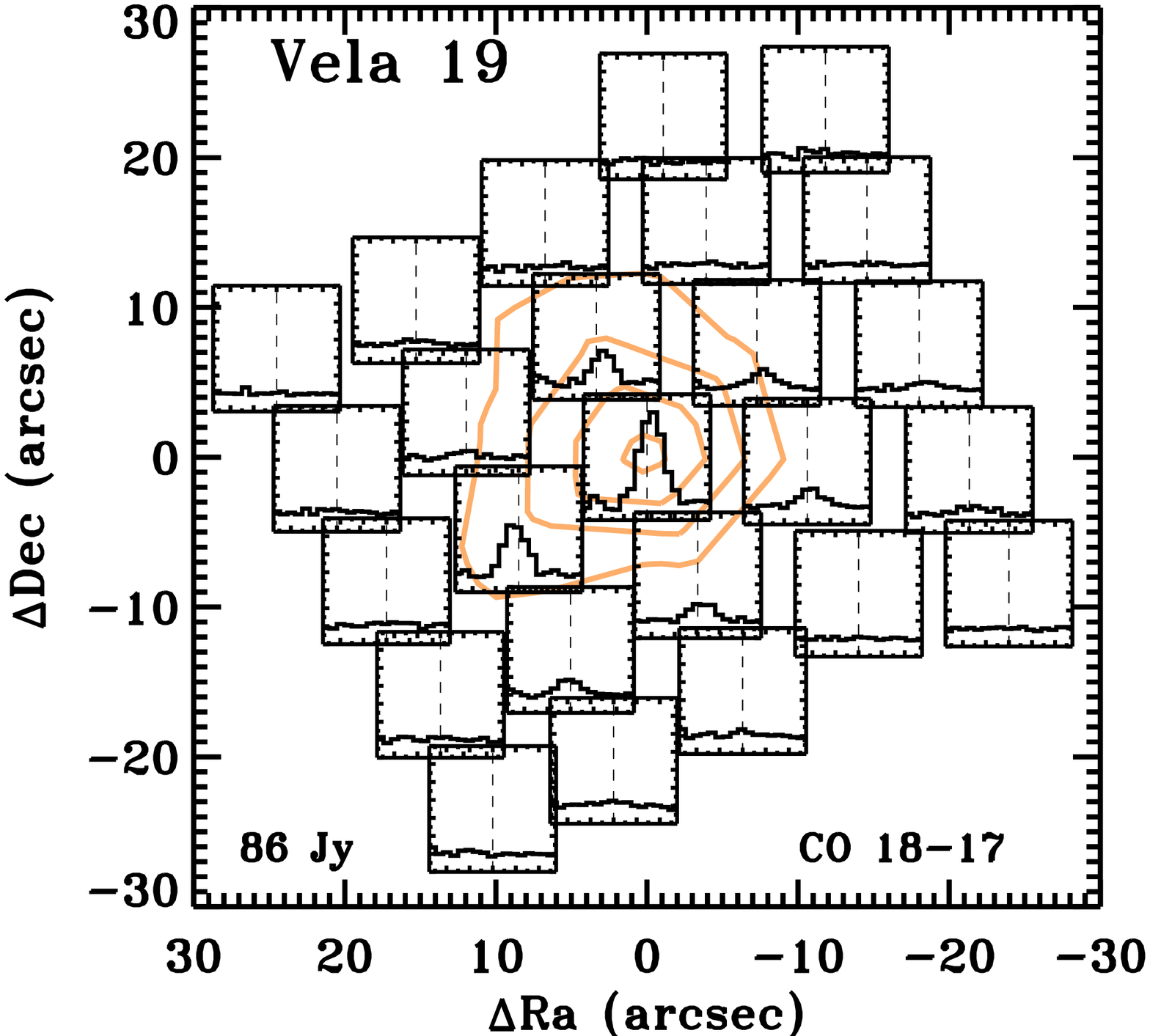}
    \end{center}
  \end{minipage}
  \hfill
  \begin{minipage}[t]{.5\textwidth}
  \begin{center}         
      \includegraphics[angle=90,height=8cm]{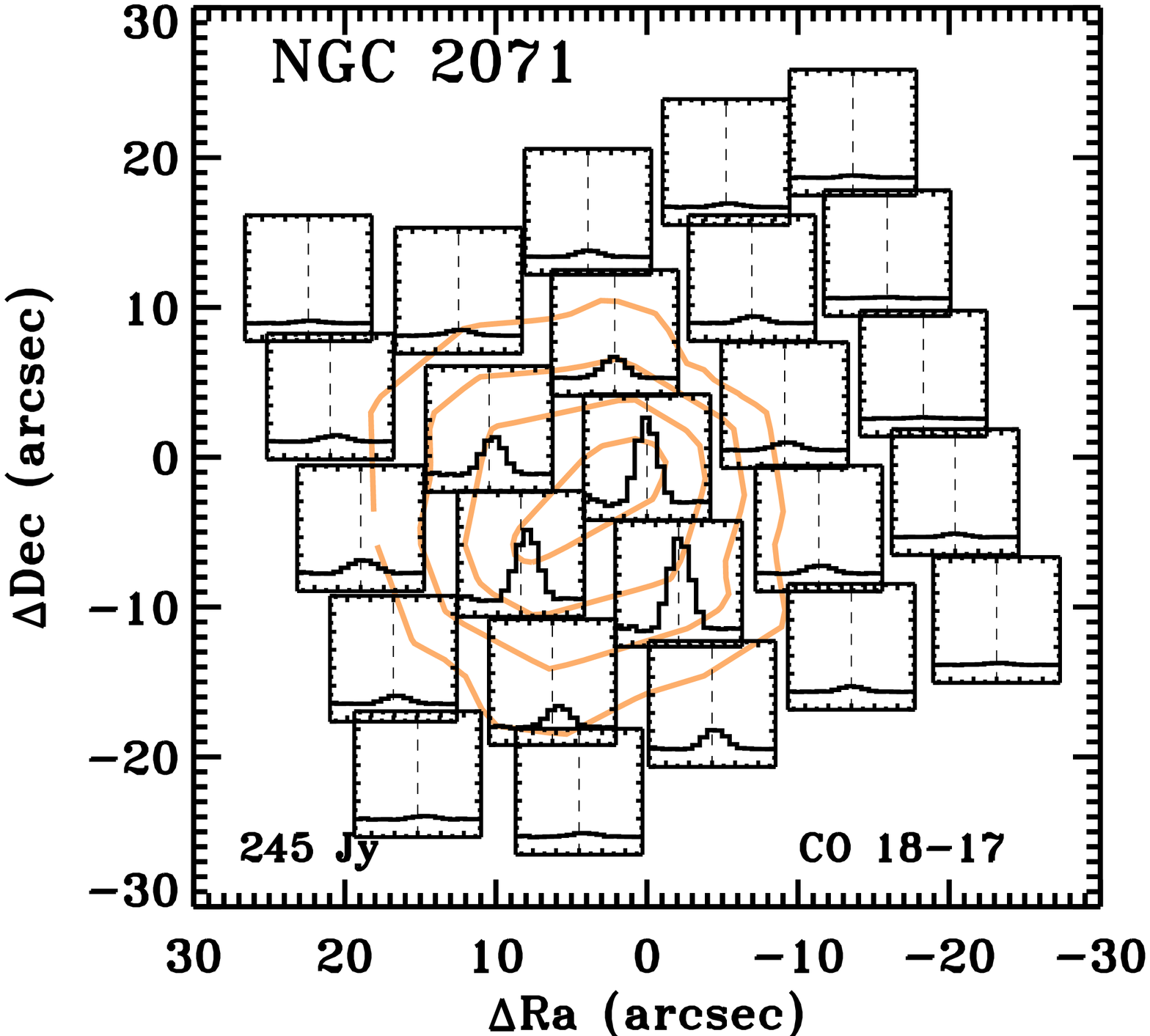}
               \vspace{+3ex}
       
    \includegraphics[angle=90,height=8cm]{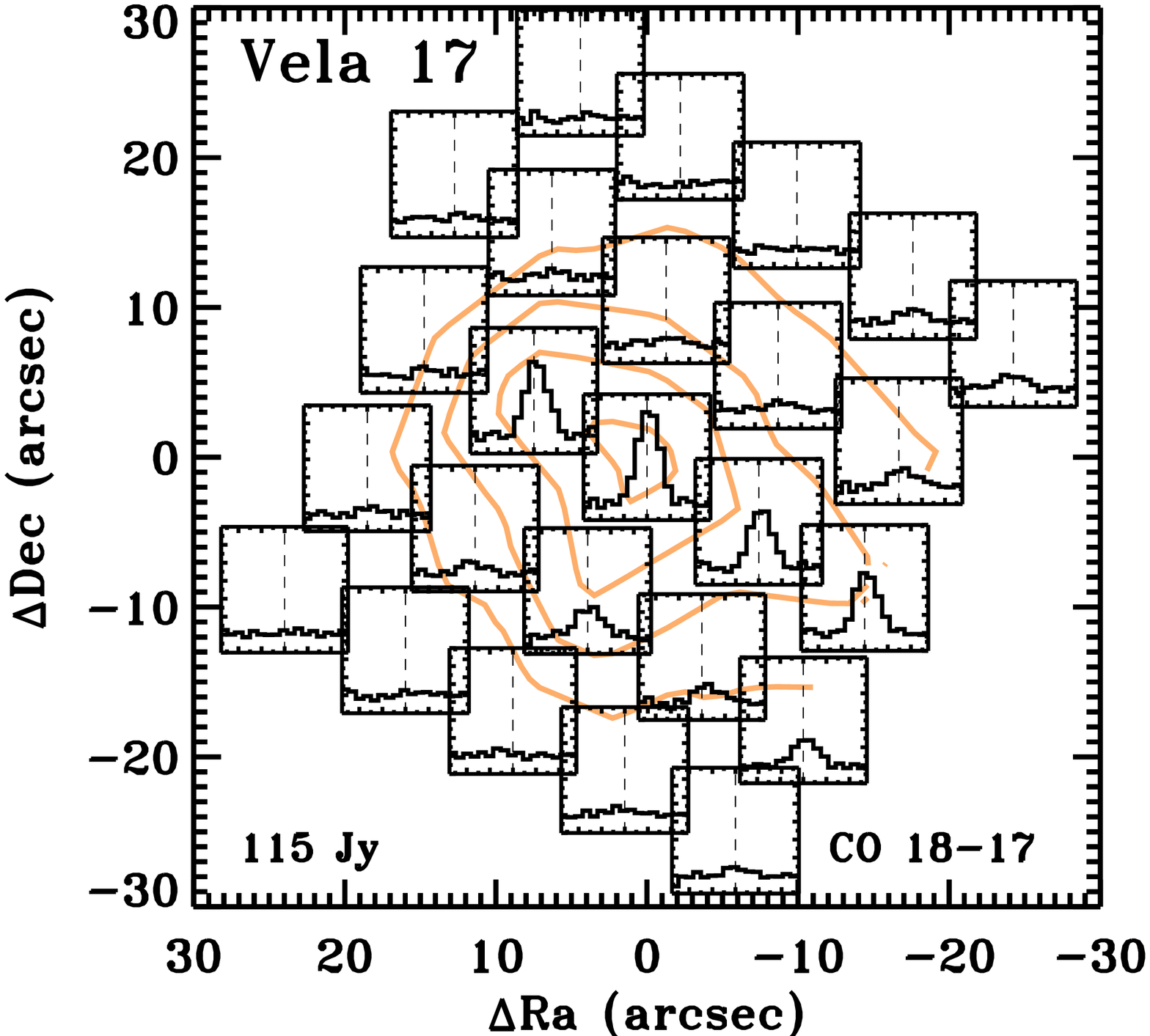}
			    \vspace{+3ex}
\includegraphics[angle=90,height=8cm]{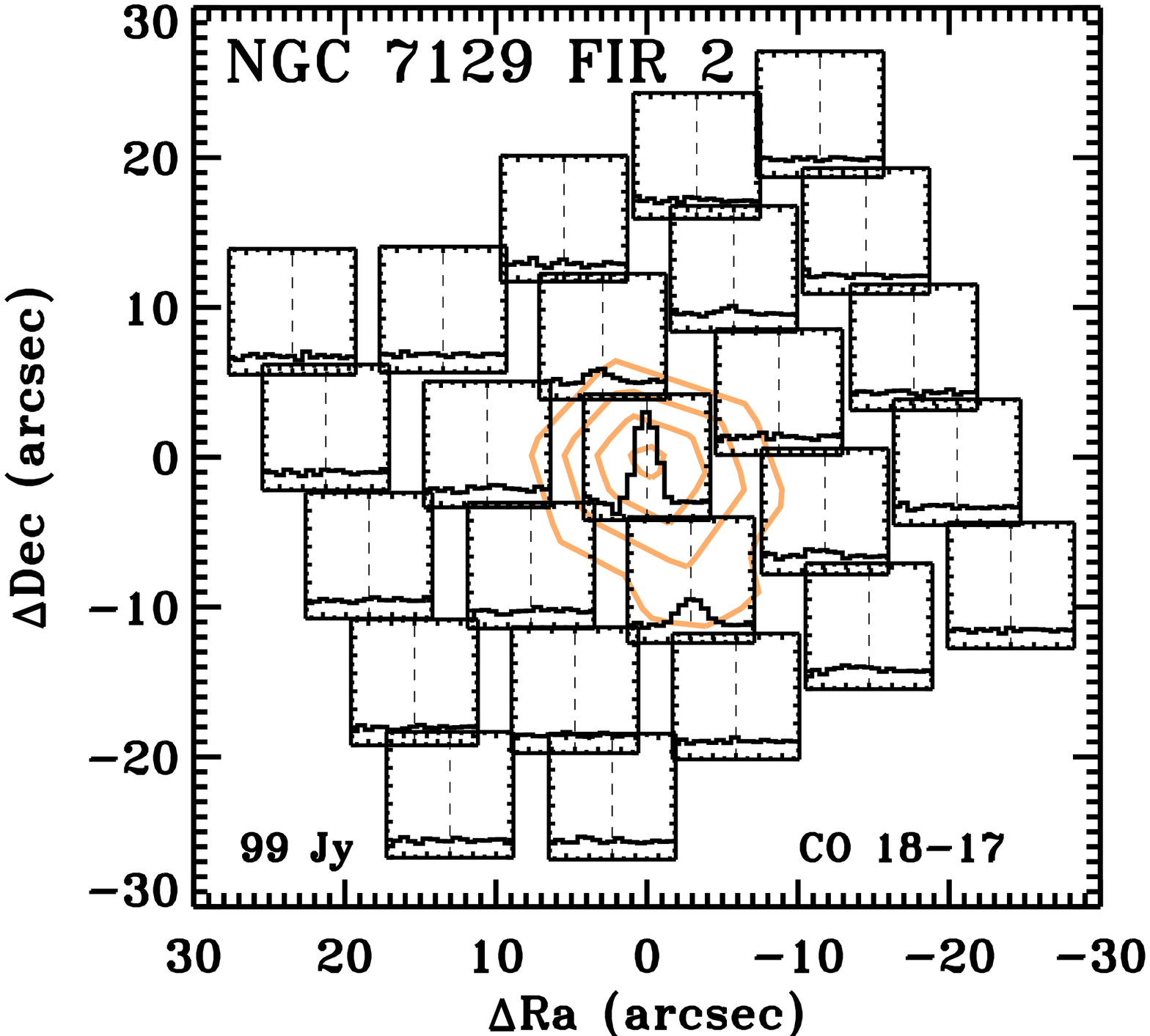}
    \end{center}
  \end{minipage}
        \caption{\label{maps} PACS spectral maps in the CO 18-17 line at 144 $\mu$m and 
        the continuum emission at 145 $\mu$m in orange contours corresponding to
         30\%, 50\%, 70\%, and 90\% of the peak value written in the bottom left corner of each map.
   		  Wavelengths in  microns are translated to the velocity 
        scale on the X-axis using laboratory wavelengths (see Table \ref{lines}) of the
species and cover the range from -600 to 600 km\,s$^{-1}$. The Y-axis shows fluxes in Jy normalized to the 
    spaxel with the brightest line on the map in a range -0.2 to 1.2.}
\end{figure*}

Table \ref{log} shows the observing log of PACS observations including
observations identifications (OBSID), observation day (OD), date of observation, total integration time, 
 and pointed coordinates (RA, DEC). Table \ref{lines} shows molecular data and observed lines fluxes 
 and upper limits for all sources that are analyzed in the paper. Figure \ref{spec} shows selected
 spectral regions to illustrate the quality of the data. Figure \ref{maps} illustrates the patterns of 
 continuum emission at 145 $\mu$m and the CO 18-17 line emission at 144 $\mu$m toward all the sources.
\begin{table*}
\caption{\label{log} Log of PACS observations}             
\centering     
\renewcommand{\footnoterule}{}  
\begin{tabular}{lccccccccccccc}     
\hline\hline       
Source & OBSID & OD & Date & Total time & RA & DEC \\
~ & ~ & ~ & ~ & (s) & ($^\mathrm{h}$ $^\mathrm{m}$ $^\mathrm{s}$) & ($^{\mathrm{o}}$ $\mathrm{'}$ $\mathrm{''}$) & ~ \\   
\hline
AFGL 490 & 1342202582 & 454 & 2010-08-11 & 3882 & 3 27 38.40  &	+58 47 08.0 \\
~		& 1342191353 & 290 & 2010-02-28 & 2029 & 3 27 38.40 & +58 47 08.0 \\
L1641 S3 MMS1		& 1342226194 & 823 & 2011-08-14 & 3882   & 5 39 55.900 & -7 30 28.00  \\
~			& 1342226195 & 823 & 2011-08-14 & 1987  & 5 39 55.900 & -7 30 28.00  \\
NGC 2071		& 1342218760 & 703 & 2011-04-17 & 3882  &  5 47 04.40 & +0 21 49.00 \\
~			& 1342218761 & 703 & 2011-04-17 & 1987  & 5 47 04.40 & +0 21 49.00  \\
Vela IRS 17		& 1342209407 & 551 & 2010-11-16 & 3882   & 8 46 34.70 & -43 54 30.5  \\
~			& 1342211844 & 594 & 2010-12-28 & 1987  & 8 46 34.70 & -43 54 30.5  \\
Vela IRS 19		& 1342210189 & 552 & 2010-11-16 & 3882   & 8 48 48.50 & -43 32 29.0  \\
~			& 1342210190 & 552 & 2010-11-16 & 1987  & 8 48 48.50 & -43 32 29.0  \\
NGC 7129 FIRS2   	& 1342186321 & 165 & 2009-10-26 & 3895   & 21 43 01.70 & +66 03 23.6   \\
~		 	& 1342186322 & 165 & 2009-10-26 & 2000  & 21 43 01.70 & +66 03 23.6   \\
\hline
\end{tabular}
\end{table*}

\onecolumn
\begin{landscape}
\begin{table}
\centering 
\caption{\label{lines} Molecular data\tablefootmark{a} and observed line fluxes}              
\renewcommand{\footnoterule}{}  
\begin{tabular}{l l c c c c c c  c  c c c c c }     
\hline\hline       
Species & Transition & Wave. & Rest Freq. & $E_\mathrm{u}$/$k_\mathrm{B}$ & $A_\mathrm{ul}$ & AFGL 490 & L 1641 & NGC 2071 & Vela 17 & Vela 19 & NGC 7129 FIRS 2 \\
~ & ~ & ($\mu$m) & (GHz) &  (K) & (s$^{-1}$) & \multicolumn{6}{c}{Fluxes (10$^{-20}$ W cm$^{-2}$)} \\
\hline    
CO & 14-13 & 185.999 & 1611.8 & 580.5 & 2.7(-4) & $11.01\pm0.11$ & $5.00\pm0.10$  & $100.13\pm0.69$  & $14.92\pm0.15$  & $7.38\pm0.10$ & $4.41\pm0.07$ \\
H$_2$O & 2$_{21}$-2$_{12}$ & 180.488 & 1661.0 & 194.1 & 3.1(-2)  & $1.62\pm0.07$ & $0.69\pm0.06$  & $12.95\pm0.09$  & $2.74\pm0.06$  & $0.74\pm0.12$ & $1.28\pm0.05$ \\       
H$_2$O & 2$_{12}$-1$_{01}$ & 179.527 & 1669.9 & 114.4 & 5.6(-2)  & absorption &$1.69\pm0.06$  & $39.81\pm0.40$  & $9.65\pm0.08$  & $2.88\pm0.06$ & $3.85\pm0.08$ \\
H$_2$O & 3$_{03}$-2$_{12}$ & 174.626 & 1716.8 & 196.8 & 5.1(-2) & $4.36\pm0.08$ &$2.25\pm0.06$  & $36.97\pm0.30$  & $6.58\pm0.07$  & $3.01\pm0.11$ & $3.13\pm0.07$ \\
CO & 16-15 & 162.812 & 1841.3 & 751.7 & 4.1(-4)& $8.84\pm0.11$ &$3.76\pm0.09$  & $85.61\pm0.45$  & $9.48\pm0.18$  & $6.32\pm0.08$ & $3.77\pm0.07$  \\
CO & 18-17 & 144.784 & 2070.6 & 945.0 & 5.7(-4) & $7.74\pm0.12$ &$3.36\pm0.13$  & $70.00\pm0.65$  & $7.62\pm0.18$  & $5.84\pm0.08$ & $3.51\pm0.13$ \\
H$_2$O & 3$_{13}$-2$_{02}$ & 138.528 & 2164.1 & 204.7 & 1.3(-1) & $2.86\pm0.08$ &$1.57\pm0.07$  & $25.60\pm0.14$  & $3.99\pm0.06$  & $1.81\pm0.06$ & $2.23\pm0.06$ \\
H$_2$O & 4$_{04}$-3$_{13}$ & 125.354 & 2391.6 & 319.5 & 1.7(-1) & $0.89\pm0.11$ &$1.80\pm0.05$  & $15.52\pm0.20$  & $2.19\pm0.10$  & $1.48\pm0.11$ & $1.57\pm0.08$ \\
CO & 22-21 & 118.581 & 2528.2 & 1397.4 & 1.0(-3) & $4.61\pm0.15$ &$2.37\pm0.11$  & $49.00\pm0.26$  & $3.68\pm0.15$  & $2.78\pm0.52$ & $2.70\pm0.11$ \\
CO & 24-23 & 108.763 & 2756.4 & 1656.5 & 1.3(-3) & $3.91\pm0.34$ &$2.25\pm0.43$  & $38.80\pm0.96$  & $3.30\pm0.29$  & $4.80\pm0.46$ & $3.23\pm0.33$  \\
H$_2$O & 2$_{21}$-1$_{10}$ & 108.073 & 2774.0 & 194.1 & 2.6(-1) & $1.47\pm0.17$ & $3.19\pm0.14$  & $35.30\pm0.30$  & $5.83\pm0.16$  & $3.12\pm0.16$ & $4.51\pm0.12$ \\
CO & 29-28 & 90.163 & 3325.0 & 2399.8 & 2.1(-3) & $<0.1$ & $1.58\pm0.57$  & $>8.17$  & $<0.46$  & $1.04\pm0.71$ & $1.58\pm0.42$  \\
H$_2$O & 3$_{22}$-2$_{11}$ & 89.988 & 3331.5 & 296.8 & 3.5(-1) & $<0.1$ & $0.99\pm0.57$  & $10.21\pm0.30$  & $<0.46$  & $<0.20$ & $1.25\pm0.42$  \\
CO & 30-29 & 87.190 & 3438.4 & 2564.9 & 2.3(-3) & $<0.15$ &  $1.58\pm0.09$  & $20.37\pm0.13$  & $2.19\pm0.19$  & $2.22\pm0.16$ & $1.58\pm0.07$ \\
H$_2$O & 7$_{16}$-7$_{07}$ & 84.767 & 3536.7 & 1013.2 & 2.1(-1) &  $<0.1$ & $<0.10$  & $<1.09$  & $<0.13$  & $<0.16$ & $<0.10$  \\
CO & 32-31 & 81.806 & 3664.7 & 2911.2 & 2.7(-3)  &  $<0.1$ & $0.92\pm0.57$  & $13.16\pm0.40$  & $<0.24$  & $<0.21$ & $0.96\pm0.29$ \\
CO & 33-32 & 79.360 & 3777.6 & 3092.5 & 3.0(-3)  &  $<0.13$ & $0.48\pm0.07$  & $12.56\pm0.24$  & $<0.10$  & $<0.16$ & $1.14\pm0.09$ \\
H$_2$O & 6$_{15}$-5$_{24}$ & 78.928 & 3798.3 & 781.1 & 4.6(-1) & $<0.17$ & $0.53\pm0.09$  & $2.71\pm0.43$  &$<0.16$  & $<0.23$ & $<0.16$ \\
CO & 36-35 & 72.843 & 4115.6 & 3668.8 & 3.6(-3) &  $<0.15$ & $<0.04$  & $7.74\pm0.21$  & $<0.18$  & $<0.36$ & $0.70\pm0.09$ \\
CO & 38-37 & 69.074 & 4340.1 & 4080.0 & 4.1(-3) &  $<0.16$ & $<0.11$  & $6.36\pm0.35$  & $<0.13$  & $<0.35$ & $<0.17$ \\
H$_2$O & 8$_{18}$-7$_{07}$ & 63.324 & 4734.3 & 1070.7 & 1.8(0) &  $<0.27$ & $0.64\pm0.11$  & $8.36\pm0.37$  & $<0.15$  & $<0.23$ & $0.57\pm0.07$ \\
\hline                  
\end{tabular}
\tablefoot{Compiled using the CDMS \citep{CDMS2,CDMS} and JPL \citep{JPL} databases. Values of Einstein $A$ coefficient 
are written in a form $A(B)\equiv A\times10^{B}$.}
\end{table}
\end{landscape}

 \end{document}